\documentclass[aps,twocolumn,floatfix,superscriptaddress,longbibliography]{revtex4-1}

\usepackage{graphicx}
\usepackage{indentfirst}
\usepackage{latexsym}
\usepackage{multirow}
\usepackage{tabls}
\usepackage{epsfig}
\usepackage{multirow}
\usepackage{subfigure}
\usepackage{comment}
\usepackage{hyperref}
\usepackage{color}
\usepackage[usenames,dvipsnames]{xcolor}
\usepackage{amssymb}
\usepackage{amsfonts}
\usepackage{amsmath}
\usepackage{bm}
\usepackage{mathrsfs}

\newcommand{\Msun}{\ensuremath{M_{ \odot }}}

\begin{document}

\title{Novel Method for Incorporating Model Uncertainties into Gravitational Wave Parameter Estimates}

\def\addCambridge{Institute of Astronomy, Madingley Road, Cambridge, CB30HA, United Kingdom}

\author{Christopher J. Moore}
\email{cjm96@ast.cam.ac.uk}
\affiliation{\addCambridge} 

\author{Jonathan R. Gair}
\email{jrg23@ast.cam.ac.uk} 
\affiliation{\addCambridge}

\date\today

\begin{abstract}
Posterior distributions on parameters computed from experimental data using Bayesian techniques are only as accurate as the models used to construct them. In many applications these models are incomplete, which both reduces the prospects of detection and leads to a systematic error in the parameter estimates. In the analysis of data from gravitational wave detectors, for example, accurate waveform templates can be computed using numerical methods, but the prohibitive cost of these simulations means this can only be done for a small handful of parameters. In this work a novel method to fold model uncertainties into data analysis is proposed; the waveform uncertainty is analytically marginalised over using with a prior distribution constructed by using Gaussian process regression to interpolate the waveform difference from a small training set of accurate templates. The method is well motivated, easy to implement, and no more computationally expensive than standard techniques. The new method is shown to perform extremely well when applied to a toy problem. While we use the application to gravitational wave data analysis to motivate and illustrate the technique, it can be applied in any context where model uncertainties exist.
\end{abstract}



\maketitle

\section{Introduction}
Bayesian techniques are widely used to draw inferences from experimental data. These rely on having a model for the signal and inferences are only as accurate as the underlying model. There are two potential problems with inaccurate models --- detection and parameter estimation. If the model is a poor representation of any signal present then the integral of the posterior over parameter space (the evidence, a common model selection metric) will be underestimated. This decreases the probability that the signal will be detected and potentially increases the chance of incorrect model selection; this is the detection problem. The parameter estimation problem is that even when a detection is made the position of the peak of the posterior evaluated using an inaccurate model may be shifted relative to the true position, possibly by much more than the statistical error arising from random noise. This shift means that there will be a systematic error in the parameter estimates. The statistical error decreases with increasing signal amplitude, but the systematic error remains constant~\cite{PhysRevD.76.104018} and is therefore most important for the loudest sources. In this work we describe how model inaccuracies can be included and marginalised over within a Bayesian framework using Gaussian process regression (GPR). This approach is straightforward to implement, incurs no additional computational cost and naturally solves both of these problems.

One situation in which model inaccuracies are known to be present is in the analysis of data from gravitational wave (GW) detectors. Over the next few years kilometre-scale laser interferometers (LIGO, Virgo) will start to take data in Advanced configurations which are expected to make detections of GWs from merging compact binaries at the rate of a few tens of events per year~\cite{2010CQGra..27q3001A}. Observations of nanohertz and millihertz GW sources using pulsar timing arrays and space-based interferometers should follow in the coming decades. These observations have the potential to transform our understanding of the astrophysics of compact objects, but only if the source parameters, e.g., the masses and spins of the compact objects, can be accurately estimated. Inference will use templates of the GWs emitted in these systems. Recent advances in numerical relativity (NR; \cite{PhysRevLett.95.121101}) have made it possible to accurately compute GW templates. These simulations are prohibitively computationally expensive, however, and inference will therefore rely on approximate templates, including post-Newtonian (PN) models (for a recent review see \cite{lrr-2014-2}), numerical ``kludge'' techniques~\cite{2007PhRvD..75b4005B} or the effective-one-body approach (see, for example, \cite{2012arXiv1212.3169D}). Use of these approximations will introduce systematic errors of the type described above and these have been shown to be potentially significant for both ground~\citep{2001PhRvD..63h2005C} and space-based detectors~\citep{PhysRevD.76.104018}. These systematic errors must be accounted for to enable correct astrophysical inference from near-future GW observations.

The best current GW models, such as EOBNR~\cite{2011PhRvD..84l4052P} or IMRPhenomC~\cite{2010PhRvD..82f4016S}, are constructed by using NR simulations to fit the values of free parameters in extended analytic models. Systematic errors are still present for these models, albeit reduced relative to pure analytic models. The approach we take to accounting for model uncertainties also exploits the fact that the model is known more precisely (although not necessarily perfectly) where NR simulations have been performed, but without needing to include arbitrary free parameters in the model.

The guiding philosophy of our approach is that of Bayesian analysis; the uncertainty in the waveform model is marginalised over, folding the information available from the NR waveforms into the prior probability distribution. This is accomplished by using GPR to interpolate the difference between the accurate waveform (here taken to be the NR waveform at the points where these are available, although using EOBNR or IMRPhenomC would also be possible) and the approximate waveform across parameter space. GPR returns a probability distribution for the waveform difference. This distribution is used as a prior to analytically marginalise the likelihood over the waveform uncertainty. The \emph{marginalised likelihood} can be used in place of the usual likelihood in any existing search algorithm. The peak of the marginalised likelihood is shifted and its width broadened to account for systematic errors, thus solving the parameter estimation problem. In addition, using the evidence associated with the marginalised likelihood as a detection statistic solves the detection problem.

In Section~\ref{sec:Theory} we demonstrate how to use GPR to construct a prior probability distribution on the waveform difference and how to marginalise the likelihood analytically over this distribution. In Section~\ref{sec:numerics} we demonstrate the efficacy of the marginalised likelihood with a toy numerical problem --- the extraction of the chirp mass of a high order PN waveform using a lower order PN waveform. We conclude with a discussion in Section~\ref{sec:discussion}.

\section{Modified likelihood}\label{sec:Theory}
Assume that the GW source is fully characterised by parameters $\vec{\lambda}$. Assume further that we have the ability to calculate the accurate (although not necessarily exact) waveform templates, $h(\vec{\lambda})$ (for notational simplicity the dependence of the templates on time is suppressed throughout). These exact templates are referred to as NR templates, and are computationally expensive to produce. Additionally, we can compute approximate waveform templates, $H(\vec{\lambda})$. These approximate waveform templates are referred to as PN templates, and are computationally cheap. The templates are related by the waveform difference, $\delta h (\vec{\lambda})$:
\begin{equation}\label{eq:waveformerror} H(\vec{\lambda})=h(\vec{\lambda})+\delta h(\vec{\lambda})\;. \end{equation}

To perform parameter estimation we must calculate the posterior distribution, $p(\vec{\lambda}|s)$, from the observed data $s$. From Bayes theorem this is proportional to the product of the likelihood, $L'(s|\vec{\lambda})$, and the prior, $\pi(\vec{\lambda})$. For a detector with stationary, Gaussian noise with power spectral density $S_n(f)$, the true likelihood is
\begin{eqnarray}\label{eq:Lexact} &L'(s|\vec{\lambda})\propto \exp\left( -\frac{1}{2} \left< s- h(\vec{\lambda})\big| s- h(\vec{\lambda}) \right> \right) \; , \\
& \textrm{where}\quad\left<a|b\right>=4\Re\left\{ \int_{0}^{\infty}\textrm{d}f\;\frac{\tilde{a}(f)\tilde{b}(f)^{*}}{S_{n}(f)} \right\} \; . \label{eq:innerprod}\end{eqnarray}
For simplicity (although it is not necessary), we will assume throughout that $\pi(\vec{\lambda})$ is flat within the relevant region ${\cal{V}}$ of parameter space. The posterior is proportional to the likelihood and the evidence becomes
\begin{equation}\label{eq:evidence} {\cal{O}}' \equiv \int L'(s|\vec{\lambda}) \pi (\vec{\lambda}) {\rm d} \vec{\lambda} \propto\int_{{\cal{V}}}\textrm{d}\vec{\lambda}\;L'(s|\vec{\lambda})\;. \end{equation}

In practice it is impossible to sample from the distribution in Eq.\ \ref{eq:Lexact} because it is prohibitively expensive to calculate the NR waveforms; instead, we must use the PN waveforms to calculate an approximate likelihood,
\begin{eqnarray}\label{eq:Lapprox} &L(s|\vec{\lambda})\propto \exp\left( -\frac{1}{2} \left< s- H(\vec{\lambda}) \big| s- H(\vec{\lambda}) \right> \right) \approx L'\; . \end{eqnarray}
The natural way to reduce the error in Eq.\ \ref{eq:Lapprox} is to construct improved (and inevitably more expensive) approximants with smaller $\delta h(\vec{\lambda})$. Instead, the proposal of this paper is to replace $L(s|\vec{\lambda})$ with a new likelihood which accounts for the uncertainty in the waveforms. The alternative likelihood is
\begin{eqnarray}\label{eq:Lcal} &{\cal{L}}(s|\vec{\lambda})\propto  \int\textrm{d}\left(\delta h(\vec{\lambda})\right)\, P(\delta h(\vec{\lambda}))\times  \nonumber\\
&\exp\left( -\frac{1}{2} \left< s- H(\vec{\lambda})+\delta h(\vec{\lambda}) \big| s- H(\vec{\lambda})+\delta h(\vec{\lambda}) \right> \right)  . \end{eqnarray}
This new likelihood has marginalised over the uncertainty in the waveform difference using the (as yet unspecified) prior $P(\delta h(\vec{\lambda}))$. The prior on the waveform difference should include the information available from the limited number of available NR waveforms and must encode our expectation that the PN waveforms are accurate at early times (or equivalently low frequencies) when the orbiting bodies are well separated, but gradually become inaccurate as the bodies inspiral. At most points in parameter space a NR waveform will not be available, and so it is necessary to interpolate the waveform difference across parameter space and simultaneously account for the error this introduces. It would seem that the problem rapidly becomes complex, and even if a suitable prior could be constructed the computational time needed to evaluate ${\cal{L}}(s|\vec{\lambda})$ would make it impractical in most contexts. Fortunately, the technique of GPR provides a natural way to interpolate the waveform differences across parameter space, incorporating all necessary prior information. GPR also has the additional property, arising from its construction, that it returns an expression for $P(\delta h(\vec{\lambda}))$ which is a Gaussian in $\delta h (\vec{\lambda})$. Since the exponential factor in Eq.\ \ref{eq:Lcal} is also Gaussian the integral may be evaluated analytically. This gives an expression for ${\cal{L}}(s|\vec{\lambda})$ which can be evaluated in the same computational time as $L(s|\vec{\lambda})$. Henceforth, for brevity, the $s$ will be suppressed in all likelihoods; e.g.\ $L(\vec{\lambda})\equiv L(s|\vec{\lambda})$.

\subsection{Gaussian process regression}\label{sec:GPR}
We will briefly summarise the key results from GPR here and refer the reader to standard texts for details~\citep{MacKay,GPR}. A useful way of thinking about the technique of GPR is as a probabilistic interpolation algorithm. Assume that we have access to NR waveforms at a few values of the parameters, $\small\{h(\vec{\lambda}_{i})|i=1,2\ldots ,n\small\}$ and can cheaply compute PN waveforms at the same parameter values. Our training set are the waveform differences
\begin{equation} {\cal{D}}=\left\{\left(\vec{\lambda}_{i},\delta h (\vec{\lambda}_{i})\right)\;|\;i=1,2,\ldots,n \right\} \; .\end{equation}
Assume that all waveform differences in the ${\cal{D}}$, and one additional difference at parameters $\vec{\lambda}$, are drawn from a multivariate Gaussian with zero mean and covariance $\bf{M}$:
\begin{equation}\label{eq:allGRPstuff} P\left(\left[\begin{matrix}\delta h(\vec{\lambda}_{i})\\\delta h(\vec{\lambda})\end{matrix}\right]\right)\sim{\cal{N}}\left({\bf{0}},{\bf{M}}\right)\;,\; {\bf{M}}=\left(\begin{matrix}{\bf{K}}&{\bf{K}}_{*}\\{\bf{K}}_{*}^{\textrm{T}}&K_{**}\end{matrix}\right), \end{equation}
where the matrix/vector/scalar
\begin{equation} \left[{\bf{K}}\right]_{ij}=k(\vec{\lambda}_{i},\vec{\lambda}_{j})\, , \, \left[{\bf{K}}_{*}\right]_{i}=k(\vec{\lambda}_{i},\vec{\lambda})\, , \, K_{**}=k(\vec{\lambda},\vec{\lambda}) , \end{equation}
are defined in terms of a covariance function, $k(\vec{\lambda}_{i},\vec{\lambda}_{j})$. Specifying the covariance function is central to GPR as it encodes our expectations about the properties of the function being interpolated. Here we use the squared exponential covariance function (sum over $a,b$)
\begin{eqnarray}\label{eq:covfunc} &k(\vec{\lambda}_{i},\vec{\lambda}_{j})=\sigma_{f}\exp\left(-\frac{1}{2}g_{ab}(\vec{\lambda}_{i}-\vec{\lambda}_{j})^{a}(\vec{\lambda}_{i}-\vec{\lambda}_{j})^{b}\right) .\end{eqnarray}
This is a very common choice of covariance function \citep{GPR} and defines a stationary, smooth Gaussian Process (GP). Other covariance functions could be considered and their applicability verified by reserving a subset of ${\cal{D}}$ to check against the interpolation. In Eq.~\ref{eq:covfunc}, a scale $\sigma_{f}$ and a (constant) metric $g_{ab}$ defining a modulus in parameter space has been defined; these \emph{hyperparameters} are determined by maximising the evidence for ${\cal{D}}$, which is discussed in Sec.\ \ref{sec:numerics}. These choices provide a simple, data-driven and quick-to-evaluate method which is ideal for parameter estimation, but the (dis)advantages of using other covariance functions should be explored in future work. If the available NR waveforms contain some uncertainty then this may be included by adding a diagonal matrix ${\bf{C}}$ to Eq.\ \ref{eq:covfunc}, where the element $C_{ii}$ is the fractional uncertainty in the NR simulation at $\vec{\lambda}_i$ times the signal-to-noise.
The choice of a zero mean GP reflects our prior belief in the validity of the PN waveforms. Our uncertainty in the PN waveforms is encoded in the ${\bf{M}}$ matrix via the choice of the covariance function. The conditional probability of the unknown waveform difference given the known differences in ${\cal{D}}$ is then
\begin{eqnarray}\label{eq:GRPprior} & P(\delta h(\vec{\lambda}))\propto \exp\left(-\frac{\left<\delta h(\vec{\lambda})-\mu(\lambda) | \delta h(\vec{\lambda})-\mu(\lambda) \right>}{2\sigma^{2}(\vec{\lambda})}\right) \; ,\\
&\mu(\vec{\lambda})= \left[{\bf{K}}_{*}\right]_{i} \left[{\bf{K}}^{-1}\right]_{ij}\delta h(\vec{\lambda}_{j}) \; , \\
&\sigma^{2}(\vec{\lambda})= K_{**}-\left[{\bf{K}}_{*}\right]_{i} \left[{\bf{K}}^{-1}\right]_{ij}\left[{\bf{K}}_{*}\right]_{j}\; .
\end{eqnarray}

\subsection{Marginalised Likelihood}\label{sec:margl}
Furnished with $P(\delta h(\vec{\lambda}))$, the marginalised likelihood may now be evaluated. The integrand in Eq.\ \ref{eq:Lcal} is the product of two Gaussians and may be evaluated analytically to give
\begin{eqnarray}\label{eq:finalresult} {\cal{L}}(\vec{\lambda})&\propto&\frac{1}{\sqrt{1+\sigma^{2}(\vec{\lambda})}}\times \\
&&\exp\left(-\frac{1}{2}\frac{\left<s-H(\vec{\lambda})+\mu(\vec{\lambda}) \big| s-H(\vec{\lambda})+\mu(\vec{\lambda})\right>}{1+\sigma^{2}(\vec{\lambda})}\right). \nonumber\end{eqnarray}
The best-fit waveform has shifted by $\mu(\vec{\lambda})$; this is the best estimate of the waveform difference returned by the GPR. The likelihood distribution is also broadened, since $\sigma^{2}(\vec{\lambda})\ge 0$. 

The likelihood in Eq.\ \ref{eq:finalresult} takes no appreciable extra time to evaluate compared to the likelihood in Eq.\ \ref{eq:Lapprox}, because the most expensive step is evaluating the inner product, which is common to both. Computing $\mu(\vec{\lambda})$ and $\sigma^2(\vec{\lambda})$ requires inverting the matrix ${\bf{K}}$; however this only depends on quantities in ${\cal{D}}$ and so may be performed offline (i.e.\ before the evaluation of the posterior begins) and saved in memory for subsequent evaluations. 

\begin{figure*}[t]
 \centering
 \includegraphics[trim=2cm 0.5cm 2cm 0.5cm, width=0.9\textwidth]{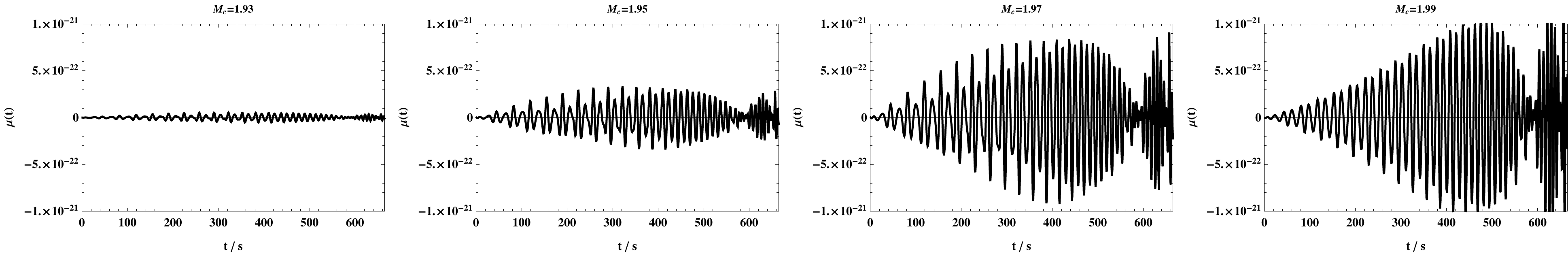}
 \caption{The mean waveform difference returned by GPR for four values of the chirp mass. The initial value of the waveform difference is zero and grows as a function of time; this is true independent of the distance in parameter space from points in ${\cal{D}}$.}
 \label{fig:mu}
\end{figure*}

\section{Numerical implementation}\label{sec:numerics}
To illustrate this method a simple search using non-spinning PN waveforms was performed. In place of the NR waveforms 3.5PN order waveforms were used, while the approximant was a 3PN waveform. A phase shift was included in the approximate waveforms to match them with the exact waveforms at the start, to illustrate the fact that GPR naturally accounts for waveform differences that are small at early times and then grow. The search was restricted to one parameter, the chirp mass ${\cal{M}}_{c}$. A signal was with intrinsic parameters $\small\{ {\cal{M}}_{c},\nu \small\}=\small\{2\,\Msun,1/5\small\}$ and extrinsic parameters $\small\{\theta_{s},\phi_{s},\iota,\psi,R,t_{c}\small\}=\small\{\pi/4,0,\pi/3,\pi/3,10\,\textrm{Mpc},0.001\,\textrm{s}\small\}$, and a short sample of the waveform between $t=-0.2\,\textrm{s}$ and $t=0$ was analysed using an analytic fit to the LIGO noise curve (from table-1 of \cite{lrr-2009-2}). For this toy problem no noise was injected into the mock data; this was so the peak of the exact likelihood was located on the true parameters and the systematic error which is the subject of this paper is clearly visible as a shift from this position. If noise was added the peak would shift by an amount consistent with the width of the posteriors but the systematic error would remain unchanged.
\begin{figure}[h]
 \centering
 \includegraphics[trim=0cm 0.5cm 0cm 0.5cm, width=0.32\textwidth]{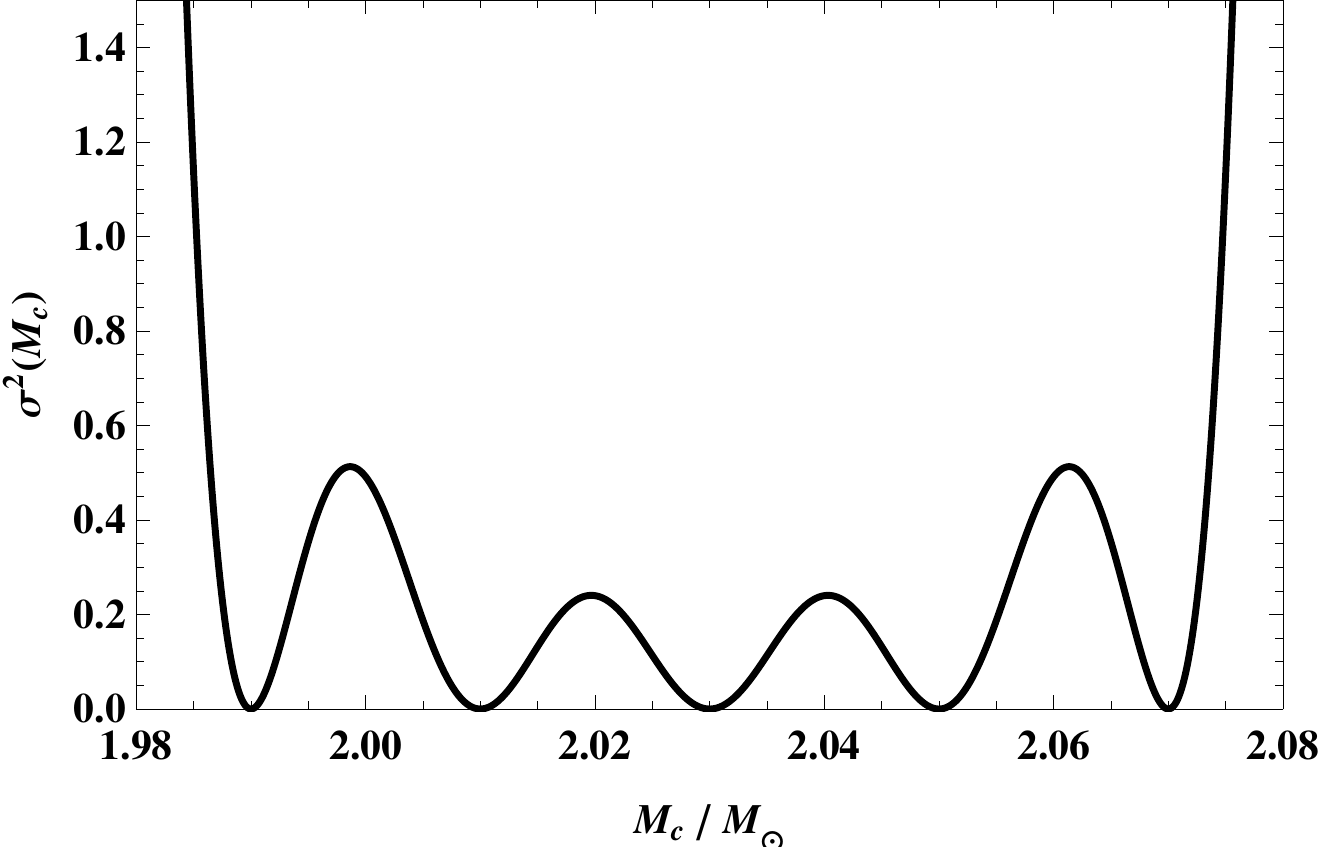}
 \caption{The GPR error estimate as a function of parameters. At parameter values in ${\cal{D}}$ the error equals zero. The GPR error does not diverge outside of ${\cal{D}}$, but the error tends rapidly to a finite limit of $\sigma_{f}$, which reflects the magnitudes of the waveform differences in the training set.}
 \label{fig:sigma}
\end{figure}

For this toy problem, ${\cal{D}}$ was taken to consist of the waveform difference computed at five chirp masses, ${\cal{M}}_{c}=\left\{1.99,2.01,2.03,2.05,2.07\right\}\times\Msun$. The matrix ${\bf{K}}$ from Eq.\ \ref{eq:allGRPstuff} was calculated an inverted offline for use during later likelihood evaluations.

The hyperparameters ($\sigma_{f}$ and $g_{ab}$) need to be chosen. This was done by maximising the evidence for ${\cal{D}}$ in Eq.\ \ref{eq:evidence} with respect to the hyperparameters \citep{MacKay,GPR}. The evidence 
\begin{equation} \label{eq:evidence} \log Z = -\frac{1}{2}\left[{\bf{K}}^{-1}\right]_{ij}\left<\delta h(\vec{\lambda}_{i}) \big| \delta h(\vec{\lambda}_{j})\right> -\frac{1}{2}\log\left|{\bf{K}}\right|\end{equation}
was maximised for $\sigma_{f}=82.0$ and $g_{ab}\small(\vec{\lambda}_{i}-\vec{\lambda}_{j}\small)^{a}\small(\vec{\lambda}_{i}-\vec{\lambda}_{j}\small)^{b}=\small({\cal{M}}_{i}-{\cal{M}}_{j}\small)^{2}/M^{2}$ where $M=0.0226\Msun$. Fixing the hyperparameters by maximising the evidence is a convenient heuristic, but other approaches could be considered. In general the hyperparameters could be marginalised over, treating them as nuisance parameters in a Bayesian search. Our approach has the advantage of performing all the calculation at the ``offline'' stage, adding no extra cost to the ``online'' search.

The PN approach breaks down near merger and hence there is poor agreement between the different PN orders; however, this just serves here to make the systematic errors which are the subject of this paper more apparent. Using an inspiral-merger-ringdown approximant, it would be expected that much smaller waveform differences, and hence smaller $\sigma_{f}$, would be obtained.

GPR can now be used to interpolate the waveform difference across the range of ${\cal{M}}_{c}$ used in the search; plotted in Fig.\ \ref{fig:mu} are examples of the interpolated waveform difference, and plotted in Fig.\ \ref{fig:sigma} is the error in the GPR analysis as a function of ${\cal{M}}_{c}$. It can be seen from Fig.\ \ref{fig:mu} that well outside ${\cal{D}}$ the GPR returns the zero function for $\mu (\vec{\lambda})$, whilst within the range of ${\cal{D}}$ it returns a smooth interpolation of the waveform difference. From Fig.\ \ref{fig:sigma} it can be seen that the GPR returns a large error outside of ${\cal{D}}$ and a small error near a point in ${\cal{D}}$.

\begin{figure*}[t]
 \centering
 \includegraphics[trim=2.2cm 0.5cm 1.6cm 0.2cm, width=0.9\textwidth]{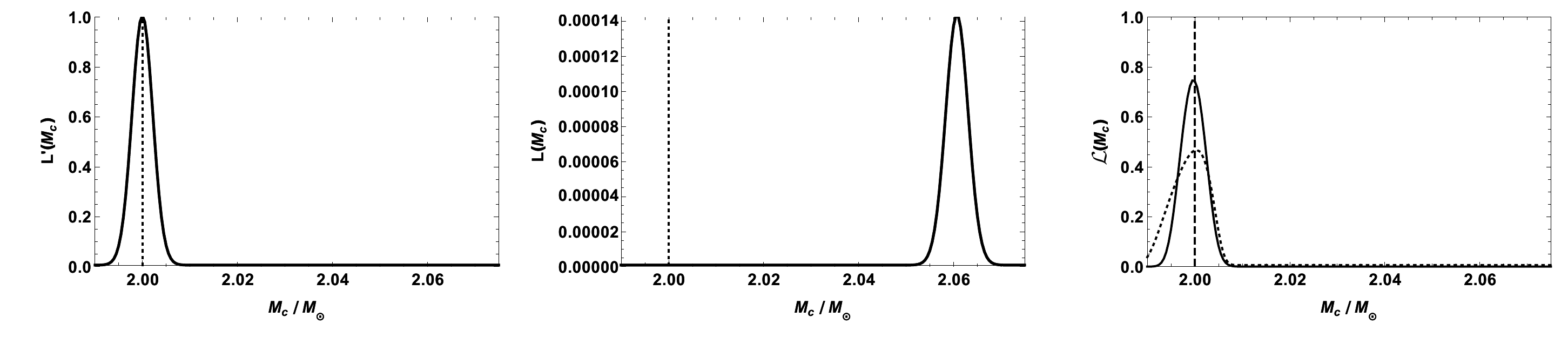}
 \caption{The panels show the likelihoods $L'(\vec{\lambda})$, $L(\vec{\lambda})$ and ${\cal{L}}(\vec{\lambda})$. The exact likelihood is peaked at the true value with value $1$ there. The approximate likelihood shows a suppressed and shifted peak. The marginalised likelihood (the solid black curve in the right-hand panel) does an excellent job of approximating the exact likelihood. The dotted curve in the right-hand panel shows the marginalised likelihood obtained using the less densely-sampled training set, ${\cal{D}}=\left\{1.984, 2.01, 2.036, 2.062, 2.088\right\}$. In this case the marginalised likelihood is broader and shorter, but still a good approximation to the exact likelihood.}
 \label{fig:LikeCompare}
\end{figure*}

Since we were working in only one dimension, the search was performed using a template grid. The resulting likelihood surfaces for $L'(\vec{\lambda})$, $L(\vec{\lambda})$ and ${\cal{L}}(\vec{\lambda})$ are shown in Fig.\ \ref{fig:LikeCompare}. The left-hand panel of Fig.\ \ref{fig:LikeCompare} was the most computationally expensive to produce due to the higher order PN waveforms; as anticipated in Sec. \ref{sec:margl}, there was no measurable difference in computational time between the centre and right-hand panels. Fig.\ \ref{fig:LikeCompare} shows that the approximate likelihood is shifted substantially with respect to the exact likelihood (the parameter estimation problem) and the evidence is dramatically reduced (the detection problem). The marginalised likelihood in the right-hand panel has addressed both of these issues.

The method outlined here does not assume that the waveform differences, $\delta h (\vec{\lambda})$, are small. In addition, although we have ignored errors in the training set waveforms for our example, the method can account for errors in the accurate waveforms. We expect the best performance when both the waveform difference and the errors in the accurate waveforms are small. However, even if this were not the case we would still expect the modified likelihood to outperform the approximate likelihood.

\section{Discussion}\label{sec:discussion}
In this paper, a novel approach to tackling the twin problems of GW detection and parameter estimation with imperfect templates is proposed. The technique involves replacing the likelihood used in standard approaches with a modified likelihood marginalised over the uncertainty in the waveform. GPR is used to interpolate the waveform difference from a training set of NR templates, and this provides the prior for the marginalisation. There have been previous attempts to improve the prospects for GW parameter estimation by interpolating accurate waveforms \cite{2013PhRvD..87l2002S}; the current approach of using GPs has the advantages of being non-parametric and automatically accounting for errors introduced by the interpolation. The resulting likelihood may be used in the same manner as the standard likelihood, and it overcomes both detection and parameter estimation problems. The new technique is exceedingly easy to implement into existing algorithms as only simple modifications to the likelihood are required. The GPR allowed us to incorporate prior beliefs about the accuracy of the templates in a non-parametric manner, and perform the marginalisation analytically. 
There is also ongoing effort into overcoming model uncertainties using parametric techniques \cite{CLPB}. The method proposed here marginalises over an unknown part of the signal to obtain the likelihood, this is similar to the marginalisation of the timing model in the context of pulsar timing array data analysis (see \cite{2014arXiv1407.1838V} for a recent treatment using GPs).

In our example the marginalised likelihood was shown to perform significantly better than the usual likelihood for both detection and parameter estimation. Crucially, the marginalised likelihood is no slower to evaluate and only a modest amount of offline calculation is required. For these reasons it is anticipated that the marginalised likelihood will be useful for future GW searches.

In this paper the focus has been on inaccuracies that arise from difficulties in building accurate models; however the method in this paper could be adapted to marginalise over any source of uncertainty in the signal assuming enough prior information is available to form a training set. Examples of errors which could be addressed in this manner include the \emph{calibration error} (a frequency dependent amplitude and phase error in the signal returned from the detector \cite{2012PhRvD..85f4034V}), the \emph{stealth bias} (a systematic error associated with a deviation from GR, before the deviation becomes detectable \cite{2013PhRvD..87j2002V}), and the error from neglecting certain physical phenomena (e.g.\ the presence of an accretion disk in a compact binary).

This approach can also be used to identify local maxima of the GPR error estimate, which could guide the NR community to regions of parameter space where new simulations are most needed. Moreover, it should have applications beyond GW data analysis. GPs are already commonly used in engineering, e.g., \cite{ko:GPBayes,Schwaighofer:GPCell}, but these techniques apply to any problem where construction of detailed models is expensive and inference relies on approximations.

The next step is to implement this technique in a higher dimensional parameter space using more realistic waveform models. This will not only provide an assessment of how significant the systematic errors from standard searches of near future advanced detector data will be, but will provide a marginalised likelihood that will give correct parameter inferences no significant additional computational cost.

\acknowledgements{We thank Christopher Berry for helpful discussions. CM is supported by the STFC. JG's work is supported by the Royal Society.}

\bibliography{bibliography}

\end{document}